**Title:**

*In Silico* Genome-Genome Hybridization Values Accurately and Precisely Predict Empirical DNA-DNA Hybridization Values for Classifying Prokaryotes

**Authors:**

Paul A. Muller Jr.*, Slava S. Epstein

* E-mail: muller.pa@husky.neu.edu

Department of Biology, Northeastern University, Boston, Massachusetts, USA.

**Abstract:**

For nearly 50 years microbiologists have been determining prokaryotic genome relatedness by means of nucleic acid reassociation kinetics. These methods, however, are technically challenging, difficult to reproduce, and - given the time and resources it takes to generate a single data-point - not cost effective. In the post genomic era, with the cost of sequencing whole prokaryotic genomes no longer a limiting factor, we believed that computationally predicting the output value from a traditional DNA-DNA hybridization experiment using pair-wise comparisons of whole genome sequences to be of value. While other computational whole-genome classification methods exist, they predict values on widely different scales than DNA-DNA hybridization, introducing yet another metric into the polyphasic approach of defining microbial species. Our goal was to develop an *in silico* BLAST based pipeline that would predict with a high level of certainty the value of the wet lab-based DNA-DNA hybridization values. Here we report on one such method that produces estimates that are both accurate and precise with respect to the DNA-DNA hybridization values they are designed to emulate.

**Keywords:**

DNA-DNA hybridization; bacteria; archaea; classification; comparative genomics

**Background:**

Classifying microorganisms into natural units is a daunting task [1], and the concept of prokaryotic species in general is a hotly debated issue [2,3]. The evolution of taxonomic thought relating to microorganisms lead to the development of the polyphasic approach [4], whereby two microorganisms are combined into the same species if they share a certain number of phenotypic and genotypic traits. A principal criterion that the query and reference strains should meet to be considered the same species relates to the degree of their gross genome similarity. DNA-DNA hybridization (DDH) is a "wet-bench" molecular biological technique that allows for the direct experimental comparison of two genomes (for a modern high throughput version of this method see Loveland-Curtze et al. [5]). This method follows simple, and seemingly indisputable, logic: the closer two microorganisms are in gene content and sequence similarity, the higher the degree their measured DNA-DNA hybridization value must be. For example, if two microorganisms share half of their genes through homology, this sets the maximum level by which they can possibly be related at the genome level to 0.5. However, homologous sequences can, and usually do, vary from each other which acts to further reduce overall relatedness (i.e. sequence dissimilarity will decrease genome relatedness from 0.5). Thus, the DNA-DNA hybridization value is a direct measure of shared gene content weighted by shared sequence similarity.

The DDH method of classification was first used by McCarthy and Bolton in 1963 [6] to establish the genome level relatedness of bacteria and now serves as the "gold standard" for novel prokaryotic species classification [4,7]. The necessary empirical evidence to justify classifying one prokaryotic species as different from another is 70% genome similarity, or less, as measured by the DDH method [4,8]. Though this method is theoretically elegant, in practice it suffers from certain limitations. These methods are laborious and require a substantial expertise, which limits their application to a few specialized labs. In the post-genomics era, one naturally wonders if there is a straightforward easy-to-implement and inexpensive *in silico* alternative that could complement, if not replace, its wet-bench counterpart using whole genome sequences.

Indeed, a number of *in silico* genome similarity measures have been advanced in the last decade for the purpose of classifying microorganisms. An early attempt was the average nucleotide identity method (ANI) developed by the Tiedji group [9] later followed by the genome-genome distance (GGD) methods summarized by Auch et al. [10]. While these methods are capable of making reasonably accurate predictions of species assignments compared to experimental DDH values from the same pair of organisms, neither method produces a metric on the same scale as DDH and therefore requires comparison against some known set of DDH values which relies heavily on prior knowledge. In addition, both of these methods take a "distance-only" approach to determining genome similarity, likely missing out on the information provided by knowledge of shared gene content. Further, due to the great variety of GGD methods to choose from, it remains unclear which method to use under what circumstances, and why.

By reasoning, the information provided by the DDH method should be easily obtained computationally by creating a composite value that combines shared gene content as well as sequence similarity. Incidentally, this should advance a more polyphasic computational approach since shared gene content not only has been shown to recapitulate phylogeny [11], independent of sequence similarity, but it also serves as an indicator of shared physiology and ecology [12,13]. Here we propose a method of directly calculating DDH from whole genome sequences, which we term *in silico* genome-genome hybridization (isGGH). This method accurately and precisely predicts experimental DDH values and serves as a useful computational method that reproduces published experimental DDH values using modern genomics tools. Because it predicts the experimental results of DDH it is directly interpretable by microbiologists. isGGH is also capable of providing convincing classification evidence at different levels of classification hierarchy from incomplete genomes by allowing the analysis of the distribution of similarity scores produced, essentially reducing the method to distance-only approaches taken by others [9,10].

We validated this method by calculating pair-wise isGGH values for a set of finished bacterial genomes previously analyzed by Goris et al. [14] using ANI and again by Auch et al. [10] using GGD methods. We then compared isGGH values to the published ANI, GGD, and empirical DDH values. We also tested the robustness of the isGGH method to undersampling by sub-sampling portions of complete genomes and then comparing the results to finished reference genomes. All species delineations herein are based on genomic information alone and we make no attempt to include true phenotypic similarity and/or ecological uniqueness. Therefore species concepts relating strictly to the phylotype and ecotype are not considered.

**Materials and methods:**

ANI values with corresponding DDH values were reported in Goris et al. [14], and were subsequently reused by Auch et al. [10] to test the GGD methods. To facilitate a three-way comparison of these methods, the same organismal genome comparisons and DDH values were used again in this study of isGGH. A total of 75 finished bacterial genomes were downloaded from the NCBI ftp website, and one from The Sanger Institute ftp website (see Supporting Table). An all-BLAST-all search was conducted on coding sequences (CDS) using the BLASTN algorithm [15], v. 2.2.25+, with default settings. Initial filtering of the BLASTN results removed all self-hits from the data set as well as all comparisons where the length of the pairwise alignment was less than 70% of the average length of the query and subject sequences. A final filter was implemented which removed all instances of multiple hits of a query sequence to a single subject genome, and *vice versa*. This final filter was necessary to reduce the great number of hits resulting from transposases, which have been shown to be the most ubiquitous of all genes, particularly in prokaryotes [16,17], greatly inflating the number of BLAST hits found for any pair of genomes in our data set.

The relatedness of an arbitrary pair of genomes is first determined by finding the number of common genes relative to the the number of CDSs of each the query genome and the reference genome, respectively:

$$\textit{fraction common CDS} = \frac{\textit{number of common CDS}}{\textit{number CDS in query genome}}, \text{ or } \frac{\textit{number of common CDS}}{\textit{number CDS in subject genome}}.$$

This provides two values in the same way DDH provides two values, depending on which is used as the test (query) and which the reference (subject). Common CDSs between any pair of species were determined from the filtered BLAST dataset. Similarity for each pair of CDSs from the BLAST dataset was determined using the H-value of Fukiya [18] (see also Ou [19]; here denoted as "h-value" so as not to confuse with Shannon's entropy H):

$$h = \frac{\frac{p_{id}}{100} * l_{aln}}{l_{query}} .$$

This h-value is the percent identity ($p_{id}$) of a pair of sequences along the entire length of the alignment ($l_{aln}$) normalized by the length of the query sequence ($l_{query}$). The percent identity, alignment length, and query length were obtained from the BLAST dataset. Since the underlying distribution of similarity scores between species is unknown and apparently asymmetrical we used the median h-value to weight the fraction of common genes shared by the two organisms to produce the isGGH values.

**Results and Discussion:**

Here, we attempted to show that the isGGH method is, at the least, comparable to other computational methods at classifying microorganisms as either the same or different species. In order to compare ANI and GGD on the DDH/isGGH scale each value was transformed linearly based on a regression equation obtained from comparison with published DDH values [10,14]. Table1 shows that the regression equations for ANI and GGD are variable from genera to genera. Because of this clade specific variability, the regression equation used was generated from the total of all points, as recommended by Auch et al [20]. In Figure 1 we show the relationship between the computed isGGH, ANI, and GGD values, after necessary linear transformations using the regression equations, and the corresponding empirically derived DDH value (ANI and DDH values taken from Goris et al. [14], GGD from Auch et al. [10]); all three computational methods explain a large portion of the variance in DDH values ($R^2$ as a measure of precision), while only isGGH has a slope near unity (a measure of accuracy) – note that without the linear transformation the slopes of ANI and GGD are 3.12 and -126.99, respectively. isGGH maintains similar statistics when the number of comparisons is increased to 71 across a more diverse set of organisms (Figure 2)[21]. When predicting species assignments corresponding to a 70% DDH cutoff, ANI and isGGH incorrectly classified six of 51 pairs of organisms and GGD misclassified eight of 51 pairs. Of the six errors made by isGGH, four were attributed to *E. coli* strain CFT073. Interestingly, its genome appears distinct enough to separate it from other strains of *E. coli* (K-12 MG1655, O157H7 Sakai, O157H7 EDL933, and O42) as a distinct species. Another possible error led to grouping *Shigella flexneri* 2a 2457T with *E. coli* K-12 MG1655 into the same species. Notably, these organisms are considered by many in the bacteriological community to be the same species [22]. Also, isGGH indicated that *Burkholderia cenocepacia* HI2424 and J2315 belong to the same species, *contra* the results from published DDH experiments [14] though in accordance with its current taxonomic standing. The ANI method overestimated the relationships between various *E. coli* strains (O127 H6 E2348 69, CFT073, K-12 MG1655, O157H7 Sakai, O157H7 EDL933, and O42) with *Shigella flexneri* 2a 2457T with respect to published DDH values. The GGD method has similar problems to ANI with various *E. coli* strains and *Shigella flexneri* 2a 2457T and as with isGGH with *E. coli* strain CFT073 and *Burkholderia cenocepacia* strains HI2424 and J2315.

We are pleased to note that the distribution of h-values (distance only measures) produced has diagnostic capabilities which are very robust to incomplete/undersampled genomes. In Figure 3 we show the shapes of the resulting distributions arising from haphazardly chosen query genomes, from various levels of classification, against a single reference genome (*Escherichia coli* K-12 MG1655). We tested how robust histogram shape is to undersampling/missing information by sub-sampling various portions of the query genome without replacement, and found that random removal of up to

95% of CDSs from the query genome had an insignificant effect on the shape of the histogram ($p = 0.09$) when compared to a complete reference genome (Figure 4). Pearson's chi-square was used to determine the consistency of shape between all histograms, following the normalization steps outlined in Porter [23].

**Conclusions:**

In 2002, an ad hoc committee for the re-evaluation of the species definition in bacteriology met in Gent, Belgium to discuss methodological developments in the field. The resulting article by Stackebrandt et al. [24] summarizing the consensus recommendations of the committee stated: "Investigators are encouraged to propose new species based upon other genomic methods or techniques provided that they can demonstrate that, within the taxa studied, there is a sufficient degree of congruence between the technique used and DNA-DNA reassociation. In addition, investigators are encouraged to develop new methods to supplement or supplant DNA-DNA reassociation." We have met this challenge by presenting a simple *in silico* method capable of predicting DNA-DNA hybridization (DDH) values accurately and precisely using whole genome sequences. Our method, isGGH, reproduces published DDH values which leads us to believe that it can also predict DDH values from previously uncompared species, thus facilitating larger-scale comparisons and shifting the focus of classification strictly to whole genome sequences. The ability of this method to accurately and precisely estimate empirical DDH values shows that it is a viable alternative to ANI and GGD, which do not mirror DDH. As a result, isGGH has the potential to be integrated seamlessly in to the field of prokaryotic classification and wherever DDH values are used.

**Acknowledgements:**

Thanks to members of the Epstein Lab for providing feedback for the manuscript. Thanks to Brittany Berdy and Maria Sizova for reviewing the manuscript for clarity. This work was performed on the Opportunity Linux cluster of Northeastern University.

**Author Contribution:**

Conceived and designed the study: PM and SE. Performed the analysis: PM. Wrote the paper: PM and SE.

**Figure Legends:**

**Figure 1. Multiple comparison of the three *in silico* genome classification methods on a common scale.**

The isGGH method (blue-squares) is compared against ANI (red-diamonds) and GGD

(yellow-triangles) for a set of 51 pairs of genomes obtained from previous studies [10,14]. To facilitate comparison on the DDH scale, which isGGH computes naturally, ANI and GGD data-points (from the NCBI BLASTN trNF 6 F3 method) were transformed using the slope and intercept from Table 1. The dashed diagonal line represents a 1:1 correspondence between the computed genome relatedness value on the x-axis and the experimentally determined relatedness value on the y-axis. A cross-hair was placed at 70% on both axes to indicate the threshold for species classification. All points above and to the right of the cross-hair are predicted to be from the same species based on both an *in silico* method and the experimental method (DDH). Points below and to the left represent pairs of genomes predicted to be from different species. Points in the upper left and bottom right are in disagreement between experimental and computational values.

**Figure 2. isGGH is both accurate and precise.**

A total of 70 comparisons where both DDH values and complete genomes were available were retrieved from previous studies (all isGGH and corresponding DDH values from Figure 1 plus 19 additional points from [21]). Accuracy of this method is determined by the slope of the regression line, and precision by the regression coefficient $R^2$. In this case the slope m is 1.06 and the coefficient $R^2$ is 0.84. Perfect accuracy would be indicated by a slope of 1.00 and precision by an $R^2$ value of 1.00.

**Figure 3. isGGH is very sensitive to genome completeness.**

Robustness of the isGGH value to genome completeness was tested for 7 genome pairs. Pairs of genomes were chosen so that their values were spread from low-relatedness to high-relatedness. For each pair, one genome was chosen as the query and the other as the subject. Portions of the query genome were randomly subsampled and compared to the complete subject genome and the isGGH value calculated. As expected there is a linear decrease in the isGGH with respect to decreasing genome completeness. From top to bottom the datapoints are: Red). *Mycobacterium tuberculosis* H37Rv against *M. bovis* AF2122; Orange). *Salmonella enterica* Ty2 against *S. enterica* LT2; Yellow). *E. coli* 536 against *E. coli* OH157H7 Sakai; Green). *Lactobacillus gasseri* against *L. johnsonii* NCC 533; Blue). *Shewanella baltica* OS155 against *S. oneidensis;* Indigo). *Pseudomonas fluorescens* Pfo1 against *P. syringeae* B728a; Violet). *Pseudomonas aeruginosa* PAO1 against *P. syringeae* tomato DC3000. The first species listed in each pair was the genome that was subsampled, while the second served as a complete reference.

**Figure 4. The distribution of h-values is extremely robust to genome completeness.** When the complete genomes of *Escherichia coli* 536 and *E. coli* O157 H7 Sakai are compared **A)** there is a distinct shape taken by the distribution of h-values with most counts being 1.0 and drastically tailing off with decreasing relation. **B)** When only 5% of the genome of *E. coli 536* is sampled and compared to *E. coli O157 H7 Sakai*, the same shape is observed ($p = 0.094$).

**Figure 5. The distribution of h-values has potential diagnostic capabilities.** Four genomes were compared to *E. coli* K12 MG1655. **A).** The distribution of h-values is shown when compared to the same species of a different strain (*E. coli* K12 W3110), **B).** a different species in the same genus (*E. fergusonii*), **C).** a different genus in the same family (*Yersinia pestis* KIM10), and **D).** a species from different family (*Shewanella baltica* OS155). As the phylogenetic relatedness decreases the distribution of h-values shifts left and the overall count of common CDS decreases.

**Tables:**

**Table 1: Statistical summaries of the three computational methods by species.**

|  | isGGH | | | ANI | | | GGD | | |
|---|---|---|---|---|---|---|---|---|---|
|  | **Rsq** | **slope** | **intercept** | **Rsq** | **slope** | **intercept** | **Rsq** | **slope** | **intercept** |
| *Burkholderia spp. (n=8)* | 0.95 | 0.82 | -8.88 | 0.96 | 2.78 | -201.44 | 0.98 | -99.56 | 86.30 |
| *Escherichia spp. (n=21)\** | 0.20 | 0.39 | 47.01 | 0.11 | 3.27 | -239.67 | 0.26 | -90.57 | 96.64 |
| *Pseudomonas spp. (n=15)* | 0.89 | 0.67 | 0.75 | 0.87 | 2.74 | -193.84 | 0.90 | -77.67 | 69.62 |
| *Streptococcus spp. (n=3)* | 0.12 | 0.38 | 61.42 | 1.00 | 13.95 | -1283.04 | 0.73 | -231.83 | 122.22 |
| *Shewanella spp. (n=3)* | 0.14 | 0.18 | 18.02 | 0.24 | 0.70 | -31.14 | 0.13 | -21.92 | 35.96 |
| **ALL\*\*** | **0.89** | **1.11** | **-10.09** | **0.95** | **3.12** | **-224.71** | **0.94** | **-126.99** | **103.02** |

*Included are *Shigella spp.*
\*\* Included are a greater variety of species than are shown in this table

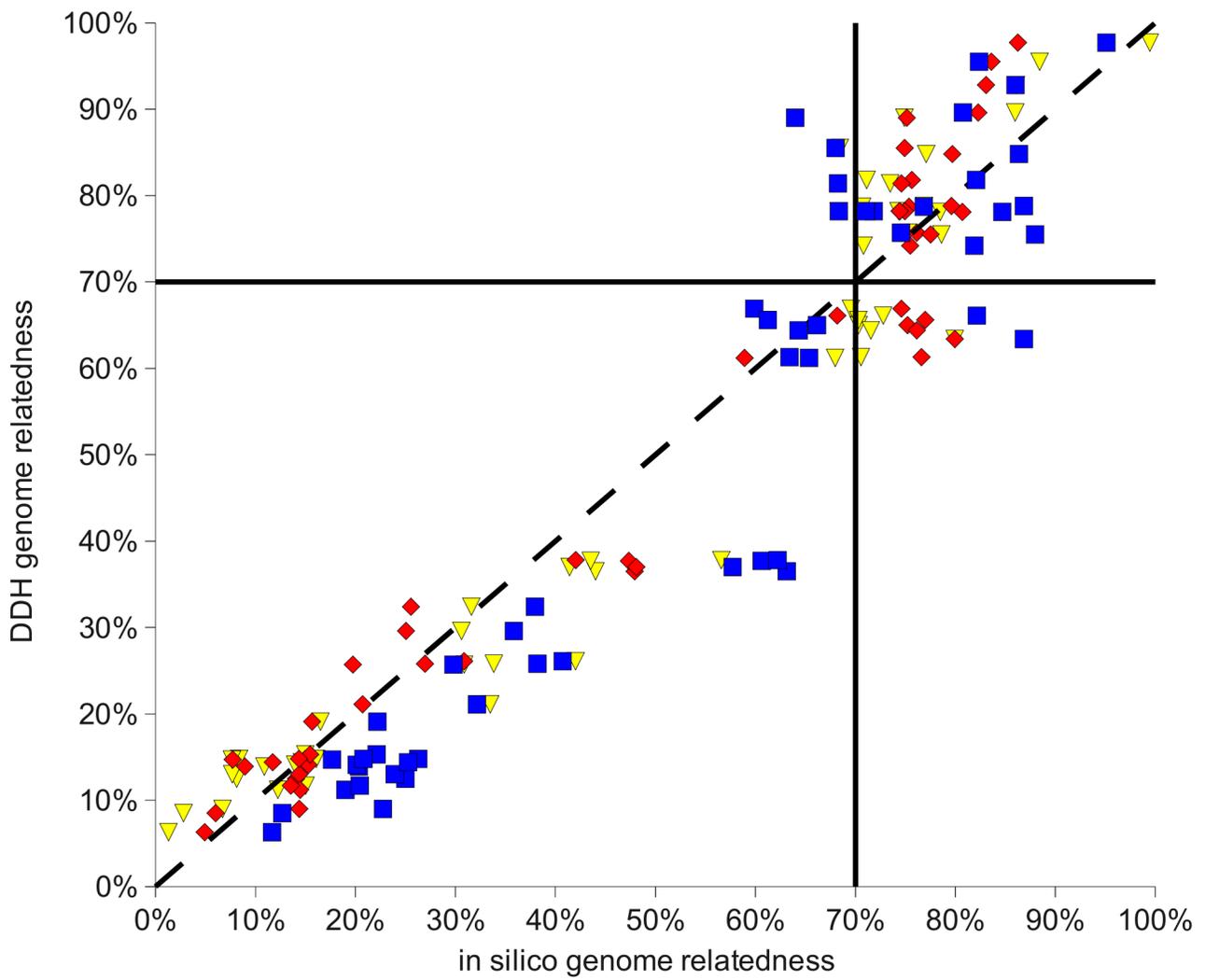

**Figure 1**

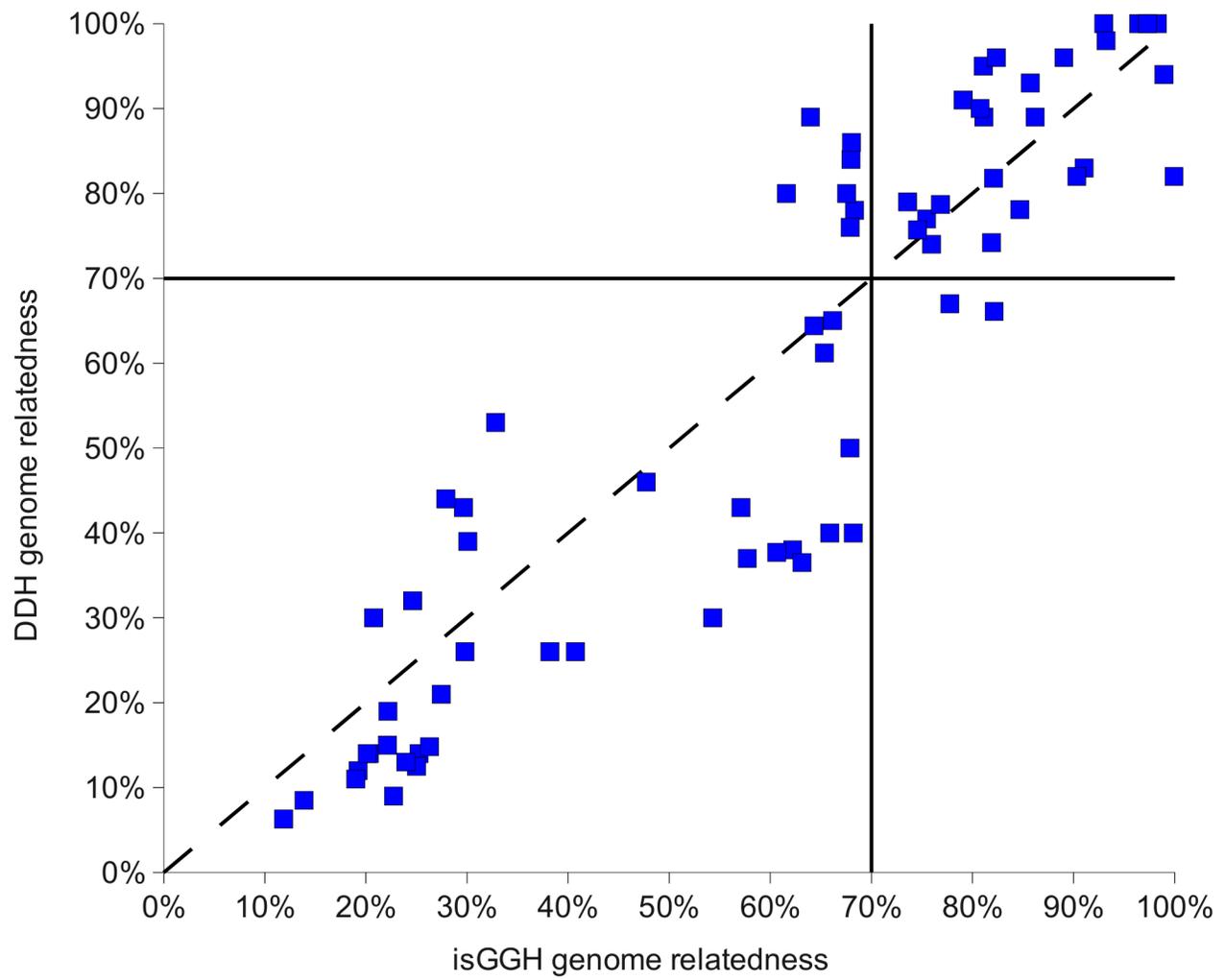

**Figure 2**

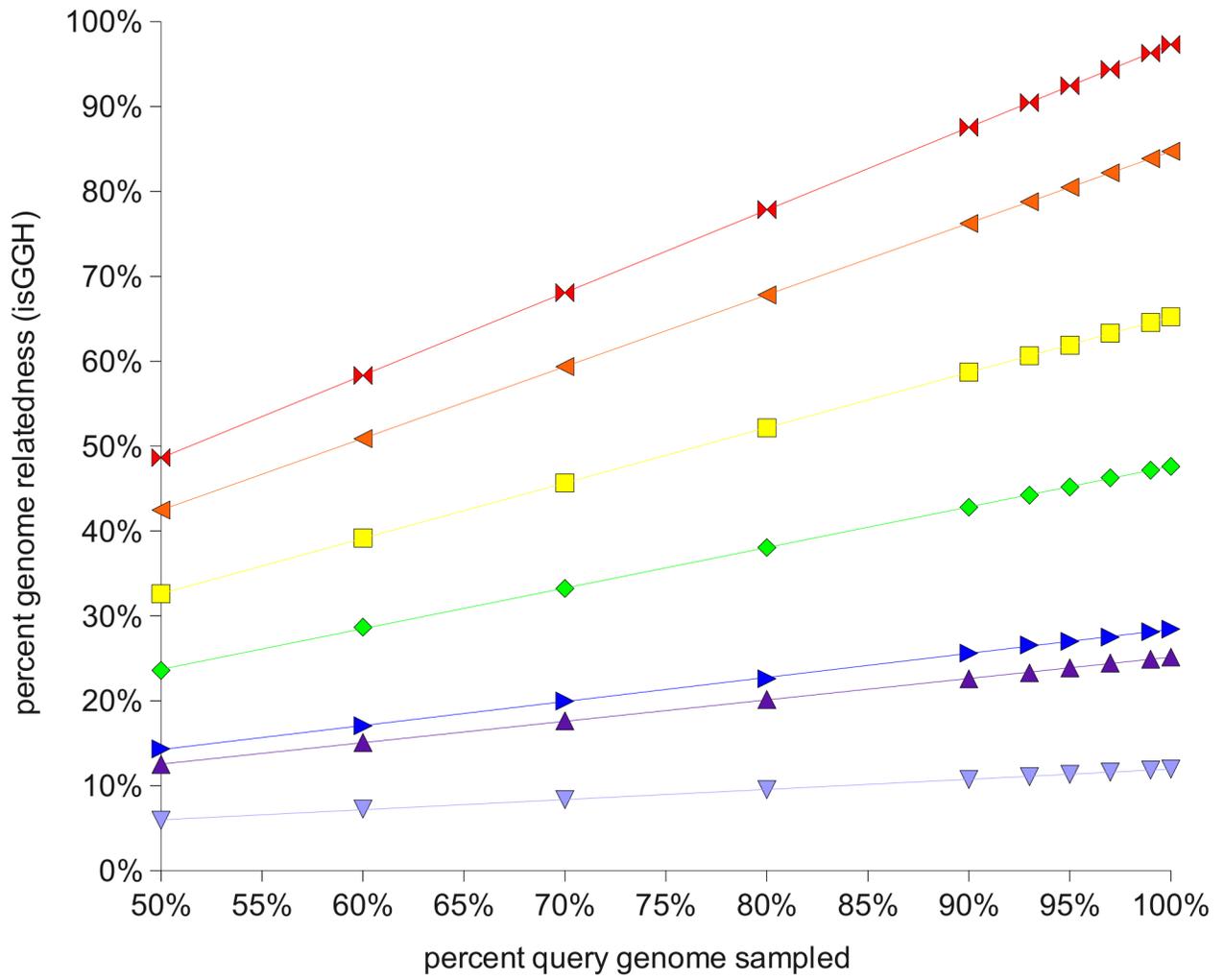

**Figure 3**

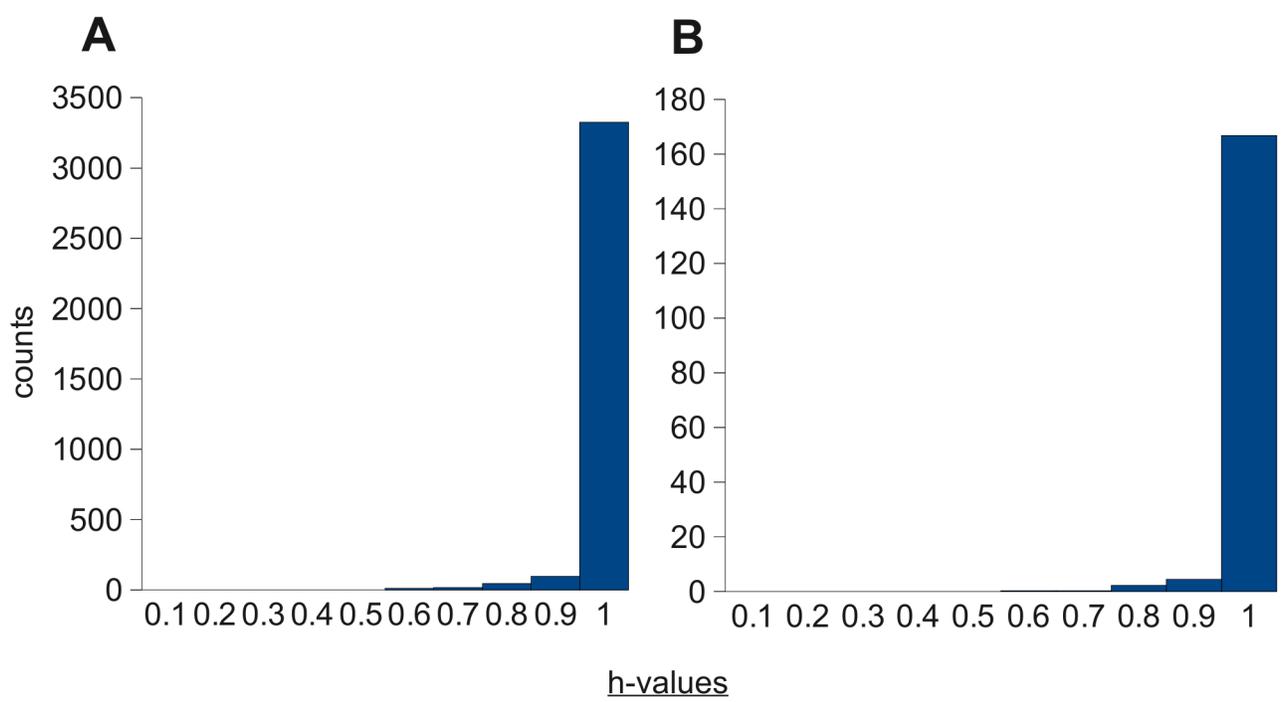

**Figure 4**

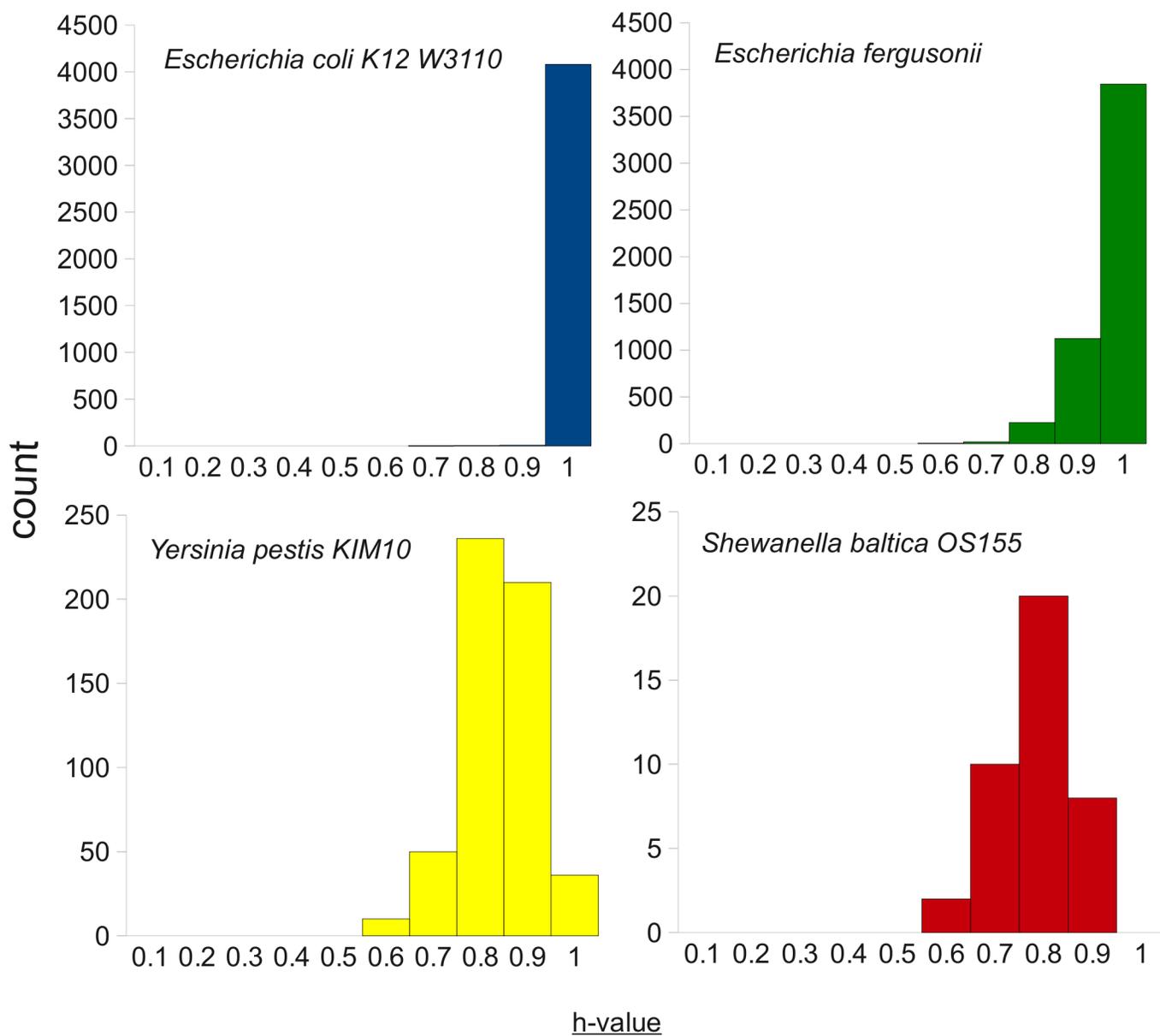

**Figure 5**

| NCBI genomes downloaded for this study | Sanger Institute genomes (ftp://ftp.s |
|---|---|
| Bacillus_anthracis_Ames_uid57909 | Escherichia_coli_O42 |
| Bacillus_cereus_ATCC_10987 | |
| Bacillus_cereus_ATCC14579 | |
| Bacillus_cereus_E33L_uid58103 | |
| Bordetella_bronchiseptica_RB50_uid57613 | |
| Bordetella_parapertussis_12822_uid57615 | |
| Bordetella_pertussis_Tohama_I_uid57617 | |
| Burkholderia_cenocepacia_AU_1054 | |
| Burkholderia_cenocepacia_HI2424 | |
| Burkholderia_cenocepacia_J2315 | |
| Burkholderia_vietnamiensis_G4 | |
| Burkholderia_xenovorans_LB400 | |
| Campylobacter_jejuni_doylei_269_97_uid58671 | |
| Campylobacter_jejuni_NCTC_11168_uid57587 | |
| Campylobacter_jejuni_RM1221_uid57899 | |
| Chlamydophila_pneumoniae_AR39_uid57809 | |
| Chlamydophila_pneumoniae_CWL029_uid57811 | |
| Chlamydophila_pneumoniae_J138_uid57829 | |
| Escherichia_coli_0127_H6_E2348_69 | |
| Escherichia_coli_536_uid58531 | |
| Escherichia_coli_CFT073_uid57915 | |
| Escherichia_coli_K_12_substr_MG1655_uid57779 | |
| Escherichia_coli_K12_substr_MG1655 | |
| Escherichia_coli_O157_H7_EDL933_uid57831 | |
| Escherichia_coli_O157_H7_Sakai_uid57781 | |
| Escherichia_coli_O157H7_EDL933 | |
| Francisella_tularensis_FSC198_uid58693 | |
| Francisella_tularensis_SCHU_S4_uid57589 | |
| Lactobacillus_gasseri_ATCC_33323_uid57687 | |
| Lactobacillus_johnsonii_NCC_533_uid58029 | |
| Mycobacterium_bovis_AF2122_97_uid57695 | |
| Mycobacterium_marinum_M_uid59423 | |
| Mycobacterium_tuberculosis_H37Rv_uid57777 | |
| Mycobacterium_ulcerans_Agy99_uid62939 | |
| Mycoplasma_capricolum_ATCC_27343_uid58525 | |
| Mycoplasma_mycoides_SC_PG1_uid58031 | |
| Pseudomonas_aeruginosa_PAO1 | |
| Pseudomonas_aeruginosa_PAO1_uid57945 | |
| Pseudomonas_fluorescens_Pf_5_uid57937 | |
| Pseudomonas_fluorescens_Pf-5 | |
| Pseudomonas_fluorescens_Pf0_1 | |
| Pseudomonas_fluorescens_Pf0_1_uid57591 | |
| Pseudomonas_fluorescens_SBW25 | |
| Pseudomonas_putida_KT2440_uid57843 | |
| Pseudomonas_syringae_B728a_uid57931 | |
| Pseudomonas_syringae_phaseolicola_1448A_uid58099 | |
| Pseudomonas_syringae_pv_B728a | |
| Pseudomonas_syringae_tomato_DC3000 | |

Pseudomonas_syringae_tomato_DC3000_uid57967
Salmonella_enterica_serovar_Typhi_Ty2_uid57973
Salmonella_enterica_serovar_Typhimurium_LT2_uid57799
Shewanella_baltica_OS155_uid58259
Shewanella_oneidensis_MR_1_uid57949
Shewanella_putrefaciens_CN_32_uid58267
Shewanella_putrefaciens_CN-32
Shigella_flexneri_2a_2457T
Shigella_flexneri_2a_301_uid62907
Streptococcus_agalactiae_2603V_R_uid57943
Streptococcus_agalactiae_A909_uid57935
Streptococcus_agalactiae_NEM316
Streptococcus_agalactiae_NEM316_uid61585
Thermus_thermophilus_HB27_uid58033
Thermus_thermophilus_HB8_uid58223
Xanthomonas_axonopodis_citri_306_uid57889
Xanthomonas_campestris_8004_uid57595
Xanthomonas_campestris_ATCC_33913_uid57887
Yersinia_enterocolitica_8081_uid57741
Yersinia_pestis_CO92_uid57621
Yersinia_pestis_KIM_10_uid57875

sanger.ac.uk/pub/pathogens/Escherichia_Shigella/)

Sheet1